\documentclass[conference]{IEEEtran}
\IEEEoverridecommandlockouts
\usepackage{cite}
\usepackage{amsmath,amssymb,amsfonts}
\usepackage{algorithmic}
\usepackage{graphicx}
\usepackage{textcomp}
\usepackage{xcolor}
\usepackage{hyperref}
\usepackage{booktabs}
\usepackage{orcidlink}
\hypersetup{pdfborder={0 0 0}}

\def\BibTeX{{\rm B\kern-.05em{\sc i\kern-.025em b}\kern-.08em
    T\kern-.1667em\lower.7ex\hbox{E}\kern-.125emX}}
\begin{document}

\title{APDT: A Digital Twin for Assessing Access Point Characteristics in a  Network \\
}

\author{
\IEEEauthorblockN{
Dharmineni Sree Yashaswinee \orcidlink{0009-0007-0440-3172} \\
\href{mailto:dharmineni.y@research.iiit.ac.in}{dharmineni.y@research.iiit.ac.in}
}
\IEEEauthorblockA{
\textit{Software Engineering Research Center} \\
\textit{IIIT Hyderabad, India}
}
\and
\IEEEauthorblockN{
Gargie Upendra Tambe \orcidlink{0009-0007-7744-9374} \\
\href{mailto:gargie.tambe@research.iiit.ac.in}{gargie.tambe@research.iiit.ac.in}
}
\IEEEauthorblockA{
\textit{Software Engineering Research Center} \\
\textit{IIIT Hyderabad, India}
}
\\

\IEEEauthorblockN{
Karthik Vaidhyanathan \orcidlink{0000-0003-2317-6175} \\
\href{mailto:karthik.vaidhyanathan@iiit.ac.in}{karthik.vaidhyanathan@iiit.ac.in}
}
\IEEEauthorblockA{
\textit{Software Engineering Research Center} \\
\textit{IIIT Hyderabad, India}
}
\and

\IEEEauthorblockN{
Y. Raghu Reddy \orcidlink{0000-0003-2280-5400} \\
\href{mailto:raghu.reddy@iiit.ac.in}{raghu.reddy@iiit.ac.in}
}
\IEEEauthorblockA{
\textit{Software Engineering Research Center} \\
\textit{IIIT Hyderabad, India}
}

}

\maketitle

\begin{abstract}
% This document is a model and instructions for \LaTeX.
% This and the IEEEtran.cls file define the components of your paper [title, text, heads, etc.]. *CRITICAL: Do Not Use Symbols, Special Characters, Footnotes, 
% or Math in Paper Title or Abstract.
Digital twins (DT) have emerged as a transformative technology, enabling real-time monitoring, simulations, and predictive maintenance across various domains, though their application in the networking domain remains underexplored.  This paper focuses on issues such as increasing client density and traffic congestion by proposing a digital twin for computer networks. Our Digital Twin, named as Access Point Digital Twin (APDT) is used for tracking user behavior and changing bandwidth demands, directly impacting network performance and Quality of Service (QoS) parameters like latency, jitter, etc. APDT captures the real-time state of networks with data from access points (APs), enabling simulation-based analyses and predictive modeling. APDT facilitates the simulation of various what-if scenarios thereby providing a better understanding of various aspects of the network characteristics. 
We tested APDT on our University network. APDT uses data collected from three access points via the Ruckus SmartZone API and incorporates NS-3 based simulations. The simulation replicates a real-time snapshot from a Ruckus access point and models metrics such as latency and inter-packet transfer time. Additionally, a forecasting model predicts traffic congestion and suggests proactive client offloading, enhancing network management and performance optimization. Preliminary results indicate that APDT can successfully predict short-term traffic surges, leading to improved QoS and reduced traffic congestion. 

\end{abstract}


\section{Introduction}
\label{intro}

% \textcolor{red}{Understanding the field and then introducing the gap} \\
% \textcolor{red}{What is unknown} \\
% \textcolor{red}{The gap that you fill} \\

Advances in Digital Twin (DT) technology have revolutionised the way complex systems are monitored, analysed, and optimised. Initially conceptualised as a virtual replica of physical systems, the DT framework has evolved to support real-time data integration and decision-making. Recent overviews, such as those by F. Tao et al. \cite{b1}, highlight the broad industrial adoption of DTs in various areas, including locomotives, healthcare, IoT systems, smart cities, and manufacturing. DTs serve as virtual representations of physical assets, enabling real-time data synchronization and predictive and, at times, prescriptive analytics. While their application in industrial environments is quite popular, deployment of DTs in wireless network management remains relatively nascent. Given the increasing complexity and scale of modern wireless infrastructures, it is important to study the utilization of DT in this domain. 

According to the Cisco Annual Internet Report \cite{b3}, the number of devices connected to IP networks was around 29.3 billion in 2023. The rapid increase in mobile devices and Internet of Things (IoT) deployments has led to dense, heterogeneous wireless environments that are inherently dynamic, resulting in volatile, hard-to-predict workloads. Traffic loads fluctuate unpredictably, interference levels vary, and network conditions are affected by numerous factors such as protocol behaviors, bandwidth allocations, and station mobility \cite{b4}. Co-channel interference, contention, and maintaining consistent quality of service (QoS) factors such as latency, jitter, throughput etc., in such environments becomes a significant challenge \cite{b5}. 

In this paper, we propose an Access Point Digital Twin (APDT), that captures the real-time operational states of wireless access points (APs), routers, and connected stations. APDT can model parameters such as client association patterns, traffic loads, interference, and hardware health indicators. APDT can be used to predict future network conditions by leveraging machine learning and data-driven analysis, enabling preemptive adjustments and optimization. The intent is to capture patterns of dynamically changing states of real twin parameters and use that data to predict or perform tasks based on the learnings. For example, when client connection patterns are low, low power-consuming states are preferred, whereas high connection activity favors states with higher throughput and lower latency.

APDT focuses on bridging the gap of i) monitoring models and system configurations, ii) using lightweight models for predictive analysis and supervising QoS iii) supporting the simulation of various scenarios that can aid in predictive maintenance of the physical counterparts. Transitioning from traditional reactive monitoring, APDT facilitates predictive analysis based on learned patterns and environmental cues. Our preliminary evaluation demonstrates the capability of APDT to forecast average byte rates over a daily cycle (section \ref{results} A) and validate these predictions through simulation, with a focus on latency metrics (section \ref{results} B). This enables the network administrators to predict future optimal configurations of network components and simulate them for enhanced confidence and fine-tuning.  As proof of concept, we implemented APDT in a university campus setting on a sub-network, utilizing historical data from three access points and simulating scenarios based on real-time configurations. The results yielded an impressive simulation accuracy of 85.03\%, alongside predictions from the predictive model, underscoring APDT’s potential for robust network optimization and management.

The rest of the paper is structured as follows: Section \ref{motivation} highlights the motivation behind the study, Section \ref{apdt} provides a description of the digital twin, Section \ref{results} presents some preliminary results of our implementation of APDT, Section \ref{relWork} discusses related work, and finally, Section \ref{conclusion} provides conclusions and possible future work. 

% \section{Background}

\section{Motivation}
\label{motivation}

% talks about heterogenity and dynamic nature
\label{ch1}
CH1: Dynamic and Heterogeneous Environments. Wireless network systems bring significant challenges such as monitoring, fault diagnosis, and system maintenance. Modern wireless environments are inherently dynamic and heterogeneous, comprising numerous components such as access points (APs), access switches, mobile client devices, and a diverse range of communication technologies (eg., 5G, 6G, Wi-Fi, Bluetooth) along with a multitude of protocols. The increasing density of parameters and the complex inter-dependencies among heterogeneous components make it hard to analyse the various parameters manually. 

% example scenario why this is necessary 
For instance, let's consider the case of a small academic institutional campus network like ours (IIIT- Hyderabad, India), consisting of approximately 100 APs, 800 mobile client devices, and multiple layers of interconnected wireless components. The tightly coupled nature of these components makes configuration and monitoring a difficult task for the network administrators responsible for the tasks. As the number and complexity of wireless elements grow, the demand for scalable and intelligent management frameworks increases. \\

% bottle necks
\label{ch2}
CH2: Bottle Neck in efficiency of decision making. For real-time analysis and decision-making, traditional centralized network management techniques are insufficient. These monitoring network systems typically rely on human-in-the-loop processing, which introduces delays in decision-making and limits scalability. Additionally, they focus on past data, which limits their ability to predict future network conditions or proactively enhance performance. Processing and real-time analysis are also hampered by the enormous amount of telemetry and log data produced by these systems. \\

% talking about client offload, band steering, channel selection
\label{ch3}
CH3: Challenges in Optimizing Access Point Configuration and Frequency Band Allocation. Each layer of the Open Systems Interconnection (OSI) model can introduce potential bottlenecks, with Layer 2 (which includes components such as access points) and below being particularly critical in terms of bandwidth limitations. Sub-networks at this level often become a primary concern for performance bottlenecks. Therefore, proactively managing client and station load balancing, as well as properly configuring channels on Layer 2 devices like access points, is essential. For example, the 5 GHz band has a greater capacity to support more stations compared to the 2.4 GHz band. As such, intelligently transferring compatible devices from lower-frequency channels (e.g., 2.4 GHz) to higher ones (e.g., 5 GHz) is crucial for maintaining a seamless user experience. To accommodate fluctuating bandwidth demands, implementing proactive and intelligent band steering mechanisms can help in managing the network more efficiently.

 % how apdt fits in and addresses these gaps 
We propose the APDT architecture to address some of these gaps by proactively maintaining, analyzing, and generating recommendations based on historical usage patterns. The overall goal of APDT is to reduce latency, enhance overall quality of service using predictive techniques, proactive optimization of client density, preemptive bandwidth allocation, and anticipatory power management.

\section{Access Point Digital Twin (APDT)}
\label{apdt}

\begin{figure}[htbp]
\centerline{\includegraphics[width=\linewidth]{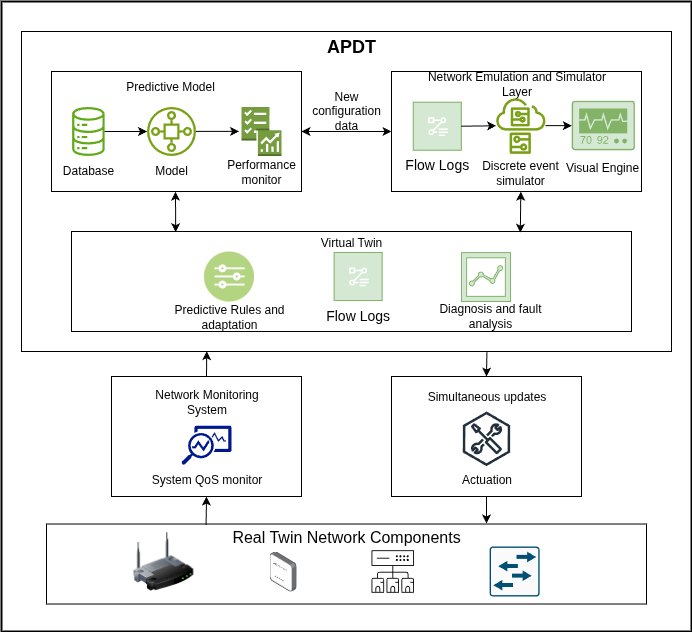}}
\caption{Architecture of the Network Digital Twin}
\label{fig1}
\end{figure}

This section details the working principles and architecture of APDT, which primarily comprises six key modules (shown in Fig. ~\ref{fig1}), namely: the real twin network, a network monitoring system, the digital twin, prediction models, a network simulator,  and an actuator. 

\subsection{Real Twin and Hardware Setup}

The "real twin" constitutes the physical foundation of the Access Point Digital Twin (APDT), encompassing the live deployment of a wireless network environment. This includes all physical infrastructure components: access points (APs), access switches, connected client devices, etc., operating in real time. In large-scale deployments such as enterprise or campus networks, this infrastructure can span multiple buildings or zones, incorporating hundreds of APs and thousands of wireless clients such as laptops, smartphones, and IoT devices. 

This component reflects the ongoing user activity, traffic load and the interaction of clients with the network. Key parameters captured include bandwidth utilization, signal strength, packet transfer rates, connection durations, and client mobility patterns, among others.  Monitoring these parameters across different network layers such as the AP, access switches, and Layer 2 or Layer 3 switches, is crucial for effective network management. 

In our implementation, the real network is represented by a subset of the university campus WLAN, comprising three Ruckus R650 dual-band 802.11ax access points installed on a single floor of an academic building. These APs are manually configured to replicate typical campus deployment settings. Operational data such as signal strength, client associations, throughput, and connection quality is collected in real time through the Ruckus SmartZone Controller API.  This setup captures the key characteristics of institute campus environments like device heterogeneity and variable traffic loads. 

The physical APs are strategically distributed to ensure comprehensive coverage of the floor and to mirror realistic operational conditions. The collected data serves as the ground truth baseline for constructing and validating the digital twin.  

\subsection{Network Monitoring System}

The network monitoring system in APDT functions as the data bridge connecting the physical components to its digital representation. Its primary purpose is to facilitate real-time, low-latency transmission of telemetry data from the real twin network to the digital twin. By facilitating continuous and structured data flow, this component allows the digital environment to maintain an accurate and time-synchronized representation of the physical infrastructure. 

This system extracts diverse performance indicators such as client association logs, signal quality metrics, channel utilization, packet loss, and throughput at granular intervals. The responsiveness of this data feed is vital to support real-time visualization, predictive modeling, and simulations within the twin. Furthermore, the network monitoring layer operates in a decoupled manner, isolating data collection from control logic, which enhances the modularity, scalability, and maintainability of the overall system.

In our implementation, this functionality is realized through the Ruckus SmartZone controller’s RESTful API \cite{b6}, which provides access to both live and historical network statistics. Periodic queries are sent to the API endpoints to gather metrics on each access point, client connections, signal parameters, and traffic statistics such as MAC addresses, connection durations, signal-to-noise ratios, and data transfer rates. The monitoring system periodically queries the SmartZone’s endpoints for metrics on each access point and associated client, such as MAC addresses, connection duration, signal-to-noise ratio, data rate, etc. The API responses are parsed, logged, and stored within a database, forming the basis for the digital twin's state synchronization.  

During development, tools like Postman \footnote{ \href{https://www.postman.com}{https://www.postman.com}}were employed to explore API endpoints, test latency, and verify data formats. The system ensures that telemetry data is transferred with minimal delay, allowing near-instantaneous updates to the digital twin’s state. By providing an automated data pipeline between the real and digital layers, the network monitoring system enables seamless communication and synchronization between the real network and its digital counterpart.  

\subsection{Digital Twin}
\label{digital Twin}

The Digital Twin serves as the digital manifestation of the real twin, replicating the real-time state and operational parameters of the physical infrastructure. It serves the purpose of connecting the network monitor with the predictive model and the simulator. Real-time gathered information: network state topology, client distribution, access point configuration, traffic conditions etc., are visualized, maintained, and monitored in this module. It has capabilities like detecting anomalies detection, performance degradation detection, and fault localization, by utilizing data stream from the network monitor. This module incorporates human intervention, suggested feedback, and actuation made by them.  

The visuals and predictive insights which are forecasted by the predictive model \ref{Predictive Models} are gathered and shown in this module. Additionally, insights and recommended configuration of both hypothetical and real-world scenarios, as simulated by the network simulator \ref{Network simulator}, are also provided here. The integrated presentation enables end users to visualize and compare outcomes from both the predictive and simulation models. This facilitates informed decision-making and adaptation across different scenarios.

In our implementation, the digital twin mirrors the state of our monitored floor — depicting AP locations, client associations, and performance metrics in real time, through the Ruckus SmartZone Controller API, allowing administrators to remotely inspect network health, trace faults, and perform historical analysis.

Incorporating intelligent systems, such as the proposed digital twin in wireless network environments, addresses the challenges faced due to a dynamic and heterogeneous environment \ref{ch1}. Primarily, it reduces the need for human intervention, thereby minimizing human errors and providing a baseline level of analytical accuracy. This is beneficial in addressing the complexities of wireless network environments. In smaller institutional setups like ours, it can be challenging for a limited IT team to manually monitor and analyze network patterns for streaming data regularly \ref{ch2}. Implementing a digital twin eases this burden by automating routine tasks, ultimately reducing operational costs and enhancing the consistency and quality of services. Moreover, automation enables faster processing and delivery, which can lead to increased efficiency and performance.

% The virtual twin functions as the core digital representation of the live wireless network in the APDT. It maintains a continuously updated, real-time visualization of the network’s state—mapping the topology, client distribution, access point configurations, traffic conditions, etc., in an interactive and   intelligible format. More than just a visual dashboard, the virtual twin incorporates diagnostic capabilities to detect anomalies, identify root causes of performance degradation, and support fault localization, which will be further elaborated in subsequent sections - Predictive Models and Network Simulator. The live telemetry from the real network offers up-to-date insights into metrics such as signal strength, traffic congestion, and client mobility, allowing in-depth analysis without interrupting the operational environment. The virtual twin also integrates logic for predictive adaptation: by learning from historical and real-time data patterns, it can trigger alerts, recommend reconfigurations, or flag emerging issues before they impact user experience. This layered functionality transforms the twin into an intelligent operations assistant rather than a passive monitor.

\subsection{Network Simulator}
\label{Network simulator}

The network simulator in APDT functions as an experimental engine that enables the modeling, testing, and visualization of both current and hypothetical network scenarios. The simulator is used to emulate “what-if” conditions that may not yet exist in the deployment but are relevant for planning or optimization. By introducing controlled variables such as increased client density, altered mobility patterns, or modified channel configurations, the simulator helps forecast how the network may behave under altered configurations. When designed with sufficient simulator fidelity, the simulator can nearly mirror the performance of the real system, providing insights into aspects of throughput, latency, interference, etc. This makes it a powerful tool for capacity planning and design validation. The simulator also plays a crucial role in the APDT, by offering a testbed to validate and refine the predictions generated by the model layer before enacting any changes on the live infrastructure.

Simulation plays a crucial role in evaluating how network configurations respond under conditions of client surge. Anticipated surges in client density (predictions made by a predictive model \ref{Predictive Models}) must be proactively addressed to ensure a seamless user experience. For instance, if a specific access point is expected to receive a significant increase in client connections, it may need to steer some clients from the 2.4 GHz band to the 5 GHz band. Such a scenario can be simulated in advance to assess its impact and potential consequences. If the tradeoffs are negligible or acceptable in light of introduced band steering, the corresponding network configurations can be adopted to effectively manage the situation. This approach not only builds confidence in the configuration decisions but also provides valuable insights into the trade-offs involved in optimizing access point configurations and frequency band allocation \ref{ch3}.

In our application, we employed NS-3, a discrete-event network simulator, to emulate network behavior under configurable conditions. To enhance simulation realism, real-time access point configurations were used to initialize the simulation environment. The simulator was then used to predict latency, achieving an approximate simulation fidelity of 73.67\% and highest being 85.03\%. Key input parameters included the number of clients, packet transmission and reception rates, and airtime utilization.  The simulator accepts configurations generated by the predictive model for validation and, in turn, produces synthetic scenario data, comprising hypothetical variations in network topology, traffic load, and user behavior. These hypothetically generated scenarios can then be used to retrain and refine the model. This bidirectional feedback loop positions the simulator as a critical component in validating, tuning, and enhancing predictive performance, thereby reinforcing the connection between model-driven insights and real-world network behavior.

\subsection{Prediction Models}
\label{Predictive Models}

The prediction models contained within the APDT architecture form the intelligence layer that enables looking into emerging network conditions. Leveraging historical telemetry data from the real network, these models can be trained to detect patterns and forecast metrics such as traffic surges, access point congestion, or signal degradation. By forecasting traffic patterns even a few intervals in advance, administrators can identify potential congestion time intervals before they occur. This prediction system enables load balancing strategies, such as offloading clients from high-traffic access points (APs) to nearby underutilized ones, in anticipation of performance degradation. Statistical models like linear regression to more advanced machine learning algorithms can be used to anticipate shifts in client behavior and network load. Once trained and validated, these models make the system transition from reactive diagnostics to proactive control, recommending configuration changes before service quality deteriorates. Crucially, the outputs of these models can be cross-validated against the network simulator, ensuring that predictions are both plausible and actionable in practice.

A non-parametric, interview-based case study conducted with the IT team of our institute revealed that APs tend to consume significant energy despite a limited number of stationary client devices are connected in a given area. The state of these APs is predominantly running in higher consumption rates despite a limited number of stationary clients. This is largely due to the fact that while Wi-Fi standards emphasize power-saving mechanisms for clients, APs are typically configured to operate at maximum capacity continuously. This results in considerable power usage. The study presented in \cite{b16} highlights methods for optimizing energy efficiency in such scenarios. Incorporating these advanced power-saving mechanisms, which include Scheduled Power Save, Dynamic Power Save, Semi-Dynamic Power Save (SDPS), Cross-Link Power Save, Wake-up Radios (WuRs), and STA offloading, can substantially reduce power consumption while maintaining, or even improving, the quality of service.

In our implementation, we collected data for two weeks from the live deployment of three Ruckus R650 access points, capturing telemetry such as client associations, signal strength, traffic volume, throughput, and packet loss. Using this dataset, we trained a lightweight linear regression model to predict traffic congestion levels at individual APs. The goal was to identify upcoming peak-load conditions and propose mitigation strategies—such as client offloading to nearby underutilized APs—before those peaks were reached. These model-generated suggestions are used for enhancing QoS, to reduce the probability of user complaints, and minimize manual intervention by network administrators. The simulation layer is then used to validate these suggested actions, providing a controlled environment to test whether proactive adjustments would lead to tangible improvements.

\subsection{Actuator}
The actuator module serves as the final stage in the APDT architecture and is responsible for applying predicted configuration changes to the live network. Based on predictions generated by the model and subsequently tested in the simulation environment, this component implements the high-level optimization decisions into the real network, such as modifying channel allocations, adjusting transmit power, or redistributing client associations. The actuator acts as a feedback loop of the digital twin and in essence, transforms APDT into a closed-loop system. It not only monitors and analyzes network behavior but also acts autonomously to improve it. This automation eliminates the need for constant manual oversight and intervention, allowing networks to adapt dynamically to fluctuating conditions in real time. The novelty of this approach lies in its full-stack integration —combining telemetry, simulation, predictive analytics, and execution, where decision-making is guided by continuous feedback and validated through simulation before being applied to the real network.

In our current prototype, execution remains a manual intervention step, used primarily for validating the relevance of predictive suggestions. For example, when a surge in client traffic is anticipated at a particular access point, our system can recommend proactively offloading specific clients to nearby APs. This could be implemented by support tools like Ruckus SmartZone API. Targeted clients can be disconnected and re-associated with alternate APs based on signal strength and load conditions. This capability shows the practical viability of a fully automated actuator, which would close the feedback loop and enable proactive optimization of wireless resources. Such integration enhances network reliability and also demonstrates a shift from reactive network management to a self-adaptive wireless environment. 

\section{Preliminary Results}
\label{results}
The experimental setup includes three access points for telemetry data collection for the predictive model, while a single AP was configured for simulation purposes. Metrics analyzed include byte rate, latency, reception time, transmission time, and airtime. For the predictive modeling component, the average byte rate was forecasted and subsequently validated using model evaluation metrics. In parallel, the NS-3 \footnote{\href{https://www.nsnam.org}{https://www.nsnam.org}} simulator employed to assess whether real-world network dynamics was effectively replicated in a digital environment. This dual evaluation aims to determine the feasibility of integrating real world dynamics into simulated settings.\\ 

\textbf{A. Model Predictions}

\begin{figure}[htbp]
\centerline{\includegraphics[width=\linewidth]{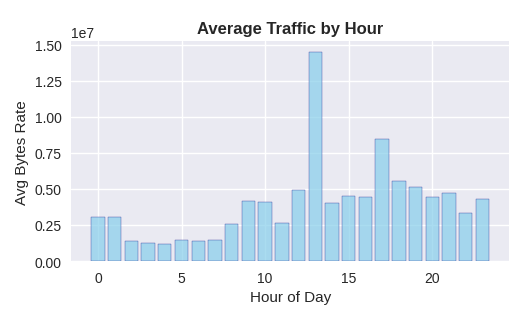}}
\caption{Average Traffic by the Hour.}
\label{fig2}
\end{figure}

\begin{figure}[htbp]
\centerline{\includegraphics[width=\linewidth]{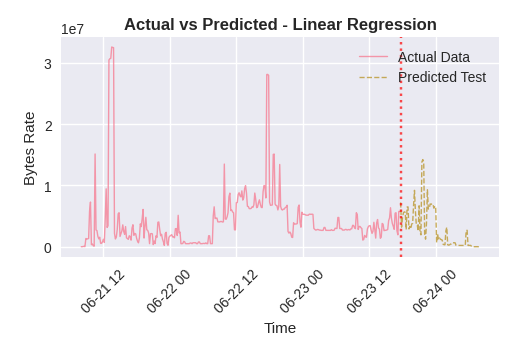}}
\caption{Predictions through Linear Regression}
\label{fig3}
\end{figure}

Fig. ~\ref{fig2} shows the hourly average byte rate. The data shows that a large number of clients were connected to the AP from noon to evening, defining a certain trend. Fig. ~\ref{fig3} visualizes a representative result from the linear regression model. It demonstrates actual and predicted traffic (with Byte Rate as the metric) over a period of several hours, highlighting the model's ability to predict the rise and fall of network load. If these predictions reach a certain threshold, suggestions can be sent to the real network to proactively offload a few clients, thereby avoiding congestion. 

Although minor deviations exist, as expected with real-world data, the predicted trend line closely corresponds to the actual traffic pattern to some extent. When deployed on a larger scale with fine-tuned models, predictive modeling can play a pivotal role in WiFi traffic forecasting systems when properly engineered and validated. \\

\textbf{B. Simulation Model}

To validate the NS3 simulator's output, real-world access point parameters, including client density, reception rate, transmission rate, airtime utilization, and noise level, were obtained from Ruckus SmartZone. A baseline assumption of 0.5 ms latency per client was incorporated. The simulation was conducted at 10-second intervals over one minute. The NS3 simulator achieved a latency prediction accuracy of 85.03\%, with an average error of 1.49 ms when compared to real-world measurements. Initial findings indicate that simulated scenarios can be highly precise when based on accurate input parameters, enabling the reliable generation of hypothetical situations. The table shown below compares real-world and simulated latency values across the simulation duration.

% Creating a table for latency comparison
\begin{table}[h]
\label{latencycomparison}
\centering
\caption{Latency Comparison: Real-World vs. Simulated}
\begin{tabular}{ccc}
\toprule
Time (s) & Simulated Latency (ms) & Real-World Latency (ms) \\
\midrule
10 & 9.4 & 14.0 \\
20 & 9.4 & 12.0 \\
30 & 9.4 & 10.5 \\
40 & 9.4 & 9.5 \\
50 & 9.4 & 9.25 \\
60 & 9.4 & 9.0 \\
\bottomrule
\end{tabular}
\end{table}

\section{Related Work}
\label{relWork}

Recent advancements in both academia and industry have explored the integration of digital twins (DTs) into wireless networking, positioning DTs as a powerful paradigm for real-time network monitoring, optimization, and predictive control.  Integration of digital twins with network infrastructure has been explored at a broad architectural level. Banisadr et al. \cite{b7} proposed the Virtual Network Twin (VNT) architecture, which combines network virtualization with digital twin concepts to create a real-time emulation of physical networks. This architecture provides enhanced network visibility, security, and robustness, offering a comprehensive framework for simulating and optimizing performance. While VNT enables high-level monitoring and mirroring, it lacks the ability to provide an autonomous feedback mechanism, a key capability addressed by our proposed APDT. 

While architecture-level solutions provide the backbone for network DTs, granular access point (AP) management is necessary for ensuring high Quality of Service (QoS). S. Kharche et al. \cite{b8} introduced a DT model specifically for Wi-Fi routers, incorporating real-time monitoring, predictive analytics, and visualizations to improve diagnostics and reduce downtime. To address access point behavior directly, L. Frank et al. \cite{b9} proposed using metaheuristic algorithms to predict user connections and dynamically optimize resource allocation. Their approach prioritizes minimizing the number of active APs. In contrast, our APDT model takes a QoS-centric approach, aiming to prevent congestion by forecasting client loads at the AP-level and adapting proactively, rather than reacting after overload occurs. 

In the commercial domain, several companies are using digital twin solutions for enterprise and university networks. Nokia, for instance, is developing DT platforms tailored for private 5G and Wi-Fi networks, enabling real-time data transfer and scenario-based simulations for planning and management \cite{b11}. Similarly, platforms such as Cisco DNA Center \cite{b12} and Aruba Central \cite{b13} incorporate machine learning (ML) driven analytics for anomaly detection and performance recommendation. While these platforms demonstrate strong monitoring capabilities, they often lack autonomous feedback loops that enable the system to act on its predictions. 

To simulate and synchronize complex wireless networks, recent research has turned to high-fidelity emulation and simulation platforms. Villa et al. \cite{b10} presented a large-scale digital twin for a 5G network, combining a physical testbed with the Colosseum emulator. This setup was used for NextG research, allowing the evaluation of experimental protocols in a controlled yet realistic setting. Along similar lines, TwiNet \cite{b14} and Colosseum both demonstrate low-latency, real-time synchronization between physical networks and their digital replicas using lightweight protocols like MQTT. These platforms highlight the growing feasibility of live, responsive digital twins in dynamic wireless environments, forming the foundation for simulations for systems like APDT. 

Beyond the wireless networking domain, digital twin frameworks have been successfully applied to smart city systems, offering design inspiration. For example, Kanigolla et al. \cite{b15} proposed WaterTwin, a digital twin designed for water quality monitoring networks. This architecture provides real-time sensing, simulation, and feedback mechanisms. The real-time analytics and predictive capabilities of WaterTwin serve as valuable reference points for our APDT framework, demonstrating how such principles can be adapted to the domain of wireless networks.

\section{Conclusion}
\label{conclusion}
This paper presented APDT, a digital twin for proactive wireless network optimization, addressing the increasing complexity of dynamically changing network parameters and the limitations of reactive management in modern deployments. By replicating real-time telemetry within a virtual model and integrating predictive analytics, our approach facilitates anticipatory decision-making and closed-loop control. We validated the feasibility of this system through a small-scale campus testbed utilizing enterprise-grade access points and live network traffic. Preliminary results showed accurate predictions of access point states and alignment with NS3 simulations, laying the groundwork for broader implementation in enterprise and campus environments. APDT introduces a promising architecture for analyzing wireless networks and proactively recommending configuration changes. Future work will aim to expand deployment, refine predictive algorithms, and assess the impact of automated actuations in production settings—moving toward the full realization of Digital Twins in wireless networking.

Our future work involves modeling the bridge between the simulation and the predictive model and evaluating the robustness of the proposed architecture. We plan to expand our scope to access layer. Along with which we plan on testing it across various scenarios and incorporate advanced algorithms for states of networking components into our predictive model.

\end{document}